# Emergent electronic insulating states in a one-dimensional moiré superlattice

Jianfeng Bi[1]†, Masaki Minamikawa[2]†, Ruige Dong[3,1]†, DongJun Kang[4]†, Zihan Weng[1], Shaoqi Sun[1], Kenji Watanabe[5], Takashi Taniguchi[6], Ryosuke Okumura[2], Huizhen Wu[1], Naoto Nakatsuji[7]*, SeokJae Yoo[4]*, Mikito Koshino[2]*, Sihan Zhao[1]*

[1] School of Physics, Zhejiang Key Laboratory of Micro-Nano Quantum Chips and Quantum Control, Zhejiang University, and State Key Laboratory of Silicon and Advanced Semiconductor Materials, Zhejiang University, Hangzhou, China
[2] Department of Physics, The University of Osaka, Toyonaka, Osaka, Japan
[3] College of Physical Science and Technology, Dalian University, Dalian 116622, China
[4] Department of Physics, Program in Semiconductor and Device, Research and Education on Next-Generation Semiconductor Materials and Devices for Chiplet Technology, and Physics Research Institute, Inha University, Incheon, Republic of Korea
[5] National Institute for Materials Science (NIMS), 1-1 Namiki, Tsukuba 305-0044, Japan Research Center for Electronic and Optical Materials, National Institute for Materials Science, Tsukuba, Japan
[6] Research Center for Materials Nanoarchitectonics, National Institute for Materials Science, Tsukuba, Japan
[7] Department of Physics and Astronomy, Stony Brook University, Stony Brook, NY, USA

† These authors contribute equally to the work.
* Correspondence: nakatsuji0323@gmail.com; seokjaeyoo@inha.ac.kr; koshino phys.sci.osaka-u.ac.jp; sihanzhao88@zju.edu.cn

**ABSTRACT**

Two-dimensional (2D) van der Waals (vdW) moiré superlattices have provided a powerful knob to engineer a plethora of new quantum states. However, extending such moiré engineering to one-dimensional (1D) vdW systems has remained challenging. Here we report the moiré-engineered electronic insulating states in a new 1D moiré superlattice, by crystallographically aligning an armchair single-walled carbon nanotube (SWNT) to 2D hexagonal boron nitride (hBN) substrate. Remarkably, we observe the emergence of pronounced insulating states at charge neutrality point (CNP), full and half moiré fillings in lattice-aligned armchair SWNT/hBN heterostructures by low-temperature electrical transport measurements. In strong contrast, armchair SWNT devices without hBN alignment do not show any of these insulating behaviors, providing compelling evidence for the significant 1D moiré effect. Our density functional theory (DFT) and tight-binding calculations reveal that synergetic nanotube partial flattening and in-plane lattice reconstruction at 1D moiré interface expand the most stable AB' stacking regions (carbon on top of boron) and open sizable band gaps at both CNP and full moiré fillings at the single-particle level. Our one-body theory predicts no band gaps at half moiré fillings, suggesting that electron correlation and/or electron-phonon interaction may give rise to these emergent insulating behaviors in our 1D moiré systems. Our work establishes a new and definite moiré engineering route for 1D vdW materials and opens an exciting avenue for exploring interaction-induced quantum phases in 1D.

*Introduction*—Moiré engineering has recently become a compelling knob to prepare and explore novel quantum states in two-dimensional (2D) van der Waals (vdW) materials. The introduction of a long-period 2D moiré superlattice via an interlayer twist and/or lattice mismatch significantly reshapes the electronic band structure and Bloch wavefunction, leading to fascinating and engineerable interaction-induced charge orderings and nontrivial topology [1–14]. However, extending such moiré protocol to engineer one-dimensional (1D) vdW systems is not straightforward. For example, in contrast to the 2D vdW counterparts, the twist between 1D constituent vdW nanotubes and/or nanoribbons is usually not controlled a priori [15–18], which makes the definite moiré engineering a challenging task for 1D vdW materials and seriously hampers the experimental study of 1D moiré physics.

Single-walled carbon nanotubes (SWNTs)-based moiré superlattices are one of the most promising testbeds to study the moiré-engineered electronic states and properties in 1D since the constituent SWNTs are mature 1D vdW materials with well-established structure-property relationship [19,20]. Previously, coaxial double-walled carbon nanotubes (DWNTs) and twisted nanoribbons formed by nanotube collapse were theoretically proposed to hold strong modulation of electronic states by moiré superlattices [21–23]. Although some progress was recently made in experimentally probing 1D SWNT-based moiré structures [24–28], experiments with a simultaneous control over moiré superlattice structures and moiré filling factors are missing. At present, there is a lack of experimental evidence for moiré-induced emergent insulating states in 1D vdW materials, and it remains unknown whether there are any physical effects that go beyond single-particle interactions in 1D moiré.

In this work, we introduce a fundamentally distinct approach to realize a novel and twist-controlled 1D moiré superlattice. Such 1D moiré is uniquely confined within an interface between two vdW materials of distinct dimensions but similar crystallographic lattice parameters—crystallographic alignment of an armchair SWNT on a hexagonal boron nitride (hBN) substrate, hereafter, lattice-aligned armchair SWNT/hBN heterostructure. Our low-temperature transport measurement discovers pronounced insulating states at full and half moiré fillings, and at charge neutrality point (CNP) in these lattice-aligned mixed-dimensional heterostructures, opening an exciting and new opportunity to explore 1D moiré physics and interactions in a well-defined 1D moiré superlattice.

*Fabrication of lattice-aligned armchair SWNT/hBN heterostructures*—We target armchair SWNTs in this study because they are truly 1D metals [29–32] and that the formed moiré lattice vector perfectly aligns with the nanotube axis at the lattice-aligned heterostructure interface. The individual armchair SWNTs in this study are noninvasively identified by the combined Rayleigh and Raman optical spectroscopies [details in Supplemental Material, Note 1] [20,33], as shown in Figs. 1(a) and 1(c) for armchair (18,18) and (22,22) SWNTs used for devices 1 and 2 in Fig. 2, respectively. The optical resonances in Figs. 1(a) and 1(c), alongside the radial breathing modes (RBMs), electronic Raman scattering (ERS) features in the corresponding Raman spectra [insets of Figs. 1(a) and 1(c)], serve as unique optical fingerprints for ambiguously identifying (18,18) and (22,22) SWNTs [20,33,34]. Nano-beam electron diffraction on additional armchair nanotubes is performed for chirality crosscheck [Supplemental Material, Fig. 1].

Figures 1(b) and 1(d) show the atomic force microscopy (AFM) images of (18,18) and (22,22)

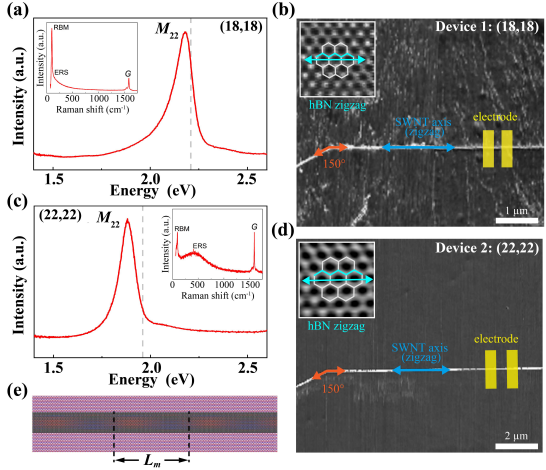

**FIG. 1. Lattice-aligned armchair SWNT/hBN heterostructures.** (a, c) Rayleigh spectra for suspended (18,18) and (22,22) SWNTs, respectively. Insets of (a) and (c): resonant Raman spectra, with excitations indicated by the grey dashed lines in (a) and (c). (b, d) AFM images of lattice-aligned (18,18) and (22,22) armchair SWNTs on hBN substrates after transfer and annealing. Insets of (b) and (d): high-resolution AFM friction images. Zigzag directions for armchair SWNTs (double blue arrows) and hBN (double cyan arrows) align. The electrodes for devices shown in Fig. 2 are overlaid. (e) Schematic of an armchair SWNT on a hBN substrate, forming a 1D moiré pattern (period $L_m$).

SWNTs on hBN on a SiO$_2$/Si substrate after the polypropylene carbonate (PPC)-assisted dry transfer and annealing processes (450 °C, < 1×10$^{-8}$ Torr) [33]. The post-annealed nanotube samples exhibit a distinctive 150° bending angle, suggesting both 0° and 30° crystallographic alignments between nanotube and hBN are preferred [35–37]. Two alignments can be clearly distinguished by resolving the atomic lattices of hBN substrate in the vicinity of nanotubes as shown by the high-resolution AFM friction images in the insets of Figs. 1(b) and 1(d). The hBN zigzag directions (cyan double arrows) clearly align with the zigzag directions of two armchair SWNT sections (blue double arrows), thereby forming lattice-aligned armchair SWNT/hBN heterostructures. A statistical analysis for post-annealed SWNT/hBN samples shows the alignment deviation is smaller than 1° [Supplemental Material, Fig. 2]. Direct experimental imaging of embedded 1D moiré interface from the curved nanotube top is challenging. We assume in this study that our lattice-aligned armchair SWNT/hBN heterostructures form a moiré period of $L_m$ ~ 14 nm [schematic in Fig. 1(e)], similar to those of the angle-aligned monolayer graphene/hBN 2D moiré superlattices [38–41] and zigzag graphene nanoribbons on hBN [42].

*Emergence of pronounced insulating states—* Figure 2(a) presents a differential conductance (*dI/dV*) map as a function of gate voltage ($V_g - V_{CNP}$) and bias voltage ($V_b$) for lattice-aligned (18,18) SWNT/hBN heterostructure (device 1). The channel length of this device is $L$ ~ 500 nm, and the data are measured at 4 K. The CNP of this device is carefully determined by examining the conductance dependence on the activated temperatures [Supplemental Material, Fig. 3]. Remarkably, we observe in Fig. 2(a) a sequence of pronounced diamond-like features on the hole-doped side that are equally spaced in $V_g$ with strongly suppressed *dI/dV* (outlined by the black dashed lines). These large diamond-like features are completely disconnected from each other, separated by $\Delta V_g$ ~ 5 V, which is qualitatively different from those arising from conventional Coulomb blockade (CB) [43–45]. The complete disconnection among these large diamonds, near-zero bias conductance profiles, and *dI/dV* magnitude inside these diamonds all indicate that the pronounced insulating states observed in Fig. 2(a) are not compatible with CB and Fabry–Pérot. The emergence of these insulating states is totally surprising since isolated armchair SWNT devices should be truly metallic. We do not observe such pronounced diamond-like insulating behaviors when the sample is electron doped [Supplemental Material, Fig. 4], which requires further studies and understanding in the future. The Supplemental Material, Fig. 5 presents a repeated *dI/dV* measurement for device 1, consistent with the data shown in Fig. 2(a) and Supplemental Material, Fig. 4.

We attribute these large diamond-like features in Fig. 2(a) to the moiré superlattice effect. To clarify which moiré fillings the insulating states correspond to, we perform high-resolution *dI/dV* measurements at regions where weaker insulating diamonds resulting from CB can be identified (although the CB effect is not as strong as a conventional semiconducting SWNT would generate). The inset of Fig. 2(a) shows the result within a small $V_g$ range between the CNP and its adjacent large diamond (2 mV stepwise in $V_g$). The observed Coulomb diamonds exhibit a single periodicity with a spacing of $\Delta V'_g$ ~ 0.06 V [inset of Fig. 2(a) and Supplemental Material, Fig. 6 for additional data], yielding a gate capacitance per unit length of $C_g$ ~ 5.3 pF/m by using $\Delta V'_g = e/(C_g \times L)$ [45–47], where $e$ is the elementary charge and $L$ ~ 500 nm is the channel length. By using the extracted $C_g$ ~ 5.3 pF/m, we obtain hole density of about 0.16 and 0.32 nm$^{-1}$ at the centers of two large diamonds next to CNP in Fig. 2(a), respectively. These densities correspond to approximately 2 and 4 holes per moiré length (~ 14 nm), which means half ($v = -2$) and full ($v = -4$) moiré fillings, respectively (four-fold degeneracy from spin and valley). We further model the device gate capacitance using COMSOL, in which we include all the key geometric and electronic parameters such as the channel length, dimensions of metal contacts, and surrounding dielectric environment, and finite density of states of a metallic SWNT. The simulation result shows a comparable $C_g$ value [Supplemental Material, Note 2 and Fig. 7].

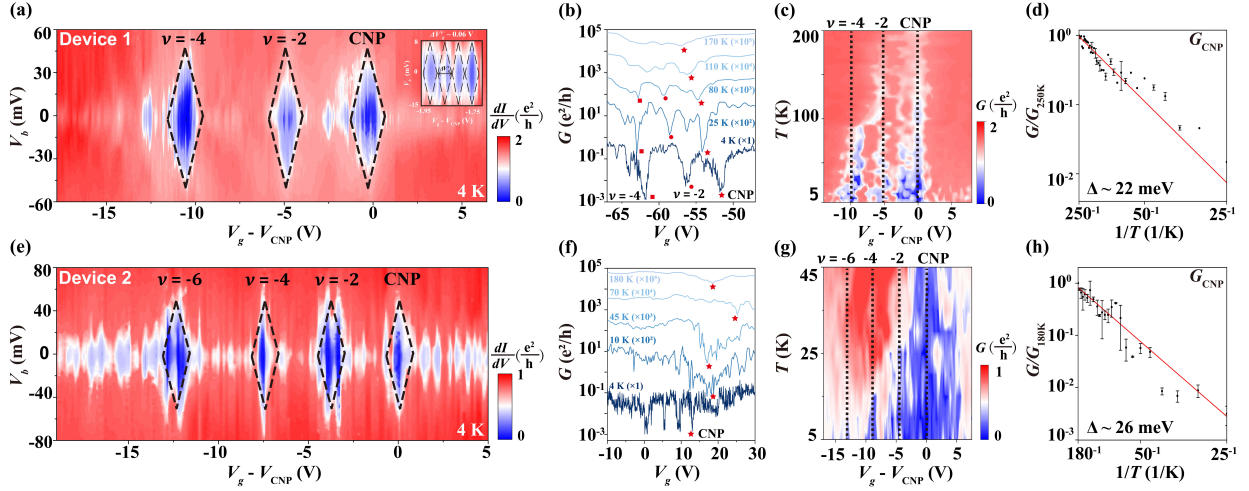

**FIG. 2. Emergent insulating states in lattice-aligned armchair SWNT/hBN heterostructures (devices 1 and 2).** (a, e) $dI/dV$ map measured at 4 K as a function of ($V_g - V_{CNP}$) and $V_b$ for (18,18) and (22,22) SWNT/hBN heterostructures, respectively. Black dashed lines highlight the emergent insulating states at CNP, $\nu = -2, -4$, etc. The separations among these pronounced diamond-like features are $\Delta V_g \sim 5$ V (a) and $\sim 4.5$ V (e), respectively. Inset of (a): high-resolution $dI/dV$ map in the vicinity of small Coulomb diamonds between CNP and $\nu = -2$ states, revealing a peak spacing of $\Delta V_g' \sim 0.06$ V and $C_g \sim 5.3$ pF/m. (b, f) Gate-dependent near zero-bias conductance measured at different temperatures for devices 1 and 2, respectively. The data are vertically offset (without horizontal offset) by different factors that are explicitly indicated, with insulating states traced by symbols. (c, g) Near zero-bias conductance as a function of ($V_g - V_{CNP}$) and temperature for devices 1 and 2, respectively. (d, h) Arrhenius plots of conductance at CNP ($G_{CNP}$) for devices 1 and 2, respectively. The red lines are fits to the thermal activation model, yielding an energy gap of $\Delta \sim 22$ and $\sim 26$ meV, respectively. The channel lengths for two devices are both $\sim 500$ nm.

The emergent insulating states at $\nu = -2$ and $-4$ can be also seen in the gate-dependent near zero-bias conductance at varying temperatures in Fig. 2(b). The stars, circles, and squares denote the states at CNP, $\nu = -2$ and $-4$, respectively. Seen from Fig. 2(b), the insulating states at CNP, $\nu = -2$ and $-4$ gradually vanish as temperature increases. The evolution of these features with temperature is presented in Fig. 2(c). Figure 2(d) shows the conductance at CNP ($G_{CNP}$) as a function of reverse temperature. We fit our data (black dots) by a thermal activation model with the form $G_{CNP}(T) \propto \exp(-\Delta/2T)$ and the fitted result is shown by the red line. The Arrhenius analysis of $G_{CNP}(T)$ yields a gap of $\Delta \sim 22$ meV at CNP. Similar analyses are performed for insulating states at $\nu = -2$ and $-4$, giving rise to gap sizes of $\sim 11$ and $\sim 13$ meV, respectively [Supplemental Material, Fig.8].

*A second heterostructure device*—We measure a second device (device 2) comprised of a lattice-aligned armchair (22,22) SWNT/hBN heterostructure and same channel length ($L \sim 500$ nm). The $dI/dV$ map of device 2 measured at 4 K is shown in Fig. 2(e). The CNP of device 2 is determined as shown in Supplemental Material, Fig. 9(a). In Fig. 2(e), we also observe a sequence of emergent insulating states (pronounced diamond-like features enclosed by the black dashed lines). The gate periodicity of these insulating states is $\Delta V_g \sim 4.5$ V, which is very close to that of device 1 ($\Delta V_g \sim 5$ V). Although we lack high-resolution $dI/dV$ data for device 2, our theoretical calculations provide support for similar $C_g$ values between the two devices [Supplemental Material, Fig. 10]. We attribute these pronounced diamonds [Fig. 2(e)] to the insulating states at CNP, $\nu = -2, -4$ and $-6$, respectively. The Supplemental Material, Fig. 9(b) repeats $dI/dV$ mapping for device 2, while the data for the electron doping is shown in Supplemental Material, Fig. 9(c). In device 2, small insulating diamond features (with nonuniform shapes and spacings) are discernible to a somewhat greater extent than in the device 1, which may be due to the slightly varying trapped states and contact resistance. The temperature evolution of the emergent insulating states for device 2 is shown in Figs. 2(f)&2(g), in which they gradually fade as temperature increases. The Arrhenius analysis of $G_{CNP}(T)$ in Fig. 2(h) yields a gap of $\Delta \sim 26$ meV at CNP, matching well the extracted gap in device 1 [Fig. 2(d)]. Comparing to device 1 [Figs. 2(b)&2(c)], the emergent insulating states of device 2, especially for $\nu = -4$, vanish at lower temperature [Figs. 2(f)&2(g)]. This slight difference is not fully understood yet.

*Armchair SWNT devices without hBN alignment*—We perform a control experiment by measuring an armchair (16,16) nanotube ($L \sim 500$ nm) on SiO$_2$/Si substrate without hBN alignment (device 3). Figure 3(a) shows its $dI/dV$ map at 4 K. In strong contrast to devices 1 and 2 with hBN alignments [Figs. 2(a)&2(e)], no pronounced diamond-like insulating states are discernible across the entire gate voltage range [Fig. 3(a)]. Even in the vicinity of CNP where the conductance is the lowest, this device exhibits high conductance larger than $e^2/h$. The gate-dependent near zero-bias conductance measurement at varying temperatures for device 3 is shown in Fig. 3(b), in which it shows an overall increasing trend in conductance as temperature decreases. Note that conductance of this device approaches $\sim 4e^2/h$ at low temperatures [Fig. 3(b)], indicating highly transparent contacts and negligibly small scattering at the channel (lower bound of mean free path $\sim 19$ μm) [48]. The temperature-dependent conductance at CNP is shown in Fig. 3(c) and it exhibits a metallic behavior as it should be [29–32], which is opposed to devices 1 and 2 [Figs. 2(d)&2(h)]. A second armchair SWNT device without hBN alignment also shows the absence of pronounced diamond-like insulating states [Supplemental Material, Fig.11]. The comparison between Figs. 2 and 3 clearly demonstrates the emergent insulating states in lattice-aligned SWNT/hBN heterostructures originate from the SWNT/hBN moiré interface.

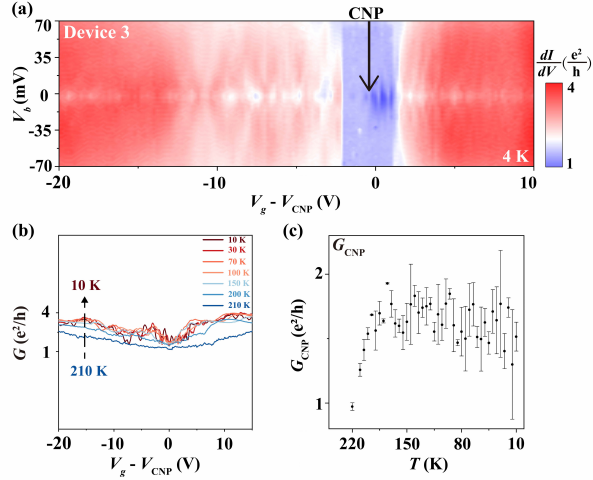

**FIG. 3. Transport measurement for an armchair (16,16) SWNT, without hBN alignment.** (a) $dI/dV$ map as a function of ($V_g - V_{CNP}$) and $V_b$ for armchair (16,16) SWNT on SiO$_2$/Si substrate measured at 4 K. (b) Gate-dependent near zero-bias conductance at various temperatures from 10 to 210 K. (c) Near zero-bias conductance measured at CNP ($G_{CNP}$) as a function of temperature. The conductance first increases as temperature decreases, and it saturates at lower temperatures, presumably due to the increased contact resistance.

*Theoretical modeling and discussion*—To shed light on moiré superlattice effect and origin of these emergent insulating features, we employ DFT and subsequent single-particle tight-binding calculations [details in Supplemental Material, Notes 3-5]. We first show the DFT-calculated commensurate SWNT aligned with hBN by assuming matched lattice constants ($a_{graphene} = a_{hBN}$) to avoid moiré scale calculation, corresponding to a simulation of AB domain (a $\sim$ 3 nm armchair SWNT used to mimic devices 1 and 2). We observe a flattened SWNT/hBN interface forms, with the most stable structure having $\varphi \sim 74°$ [top panel of Fig. 4(a); details in Supplemental Material, Note 5]. We note that the height change induced by this flattening is about 0.3 nm, which is only about 10% of the nanotube diameter, and is not noticeable by our AFM height measurement.

Figures 4(b) and 4(c) show the calculated interlayer binding energy of the flattened structure shown in top panel of Fig. 4(a) without and with lattice relaxation, respectively. The bottom panel of Fig. 4(a) schematically illustrates how the interlayer binding energy is mapped onto the 2D x-y plane for clarification, where x and y are along the tube axial and circumferential directions, respectively. Upon relaxation [Fig. 4(c)], the system expands the most stable AB' stacking domains (indicated by the yellow dashed lines), forming a 1D periodic moiré structure. Figure 4(d) illustrates the atomic stacking configurations for AB' (top panel) and saddle point (SP) (bottom panel), where carbon atoms of the SWNT are aligned directly over boron atoms of hBN in AB' and that the SP corresponds to a transitional region that is energetically unfavorable. The expansion of the most stable AB' domains is a general trend upon lattice relaxation that does not depend on the angle $\varphi$ [Supplemental Material, Fig. 12].

Figure 4(e) presents the calculated single-particle band structure of the relaxed 1D moiré structure shown in Fig. 4(c). It predicts a sizable band gap of approximately 12.1 meV at the CNP [($g_3$ in Fig. 4(e)], which is comparable to those extracted from Figs. 2(d) and 2(h) for devices 1 and 2. However, the experimentally observed gaps are still larger, which may hint at additional electron-electron interaction at work [41,49]. It also predicts the opening of a smaller gap of $\sim$ 2.5 meV at $\nu = +4$ and $\sim$ 5.3 meV at $\nu = -4$ [($g_4$ and $g_2$ in Fig. 4(e)], the latter of which approximately lies in between the two gaps for devices 1 and 2. We further calculate the evolution of the single-particle band gap with varying $\varphi$, and find that the gap is positively correlated with $\varphi$ [Supplemental Material, Fig. 13].

Our theory at the single-particle level predicts no gapped states at half moiré fillings $\nu = \pm 2$ [Fig. 4(e)], while we experimentally observe these insulating states at $v = -2$ [Figs. 2(a) and 2(e)]. The physics behind is clearly beyond the single moiré-period-modulated one-body spectrum. One possible scenario is the charge density wave (CDW) ordering with doubled moiré period due to the intrinsic electron-electron correlation or spontaneous strain (Peierls instability) [2,3,50–52], in which each of K and K' bands are independently folded into the half-sized moiré Brillouin zone (BZ), with a natural gap opening at the new BZ boundary. Within this picture, the CDW charge order may compete with the underlying moiré period modulation and the system may open a gap when the CDW and the moiré period are commensurate [6]. The moiré potential of the current system is weak, where we don't observe well-defined quantum-dot like discrete levels in our calculation [Fig. 4(c)]. This fact may make the Mott insulator less likely to be the mechanism since electrons in the system can behave far from the Hubbard picture with a tightly bound character [53]. Another possibility is the interband hybridization between K and K' states by, for example, Kekulé pattern ordering [54]. It is noticeable that angle-aligned graphene/hBN 2D moiré superlattices do not show the gapped states at half moiré fillings [38–41]. This is a signature of strong interactions that 1D has. Although the underlying physical mechanism at half moiré filling has not been fully elucidated, our results are expected to serve as a catalyst for increased theoretical and investigative efforts in moiré superlattices and interaction-induced charge orders in 1D [55–59].

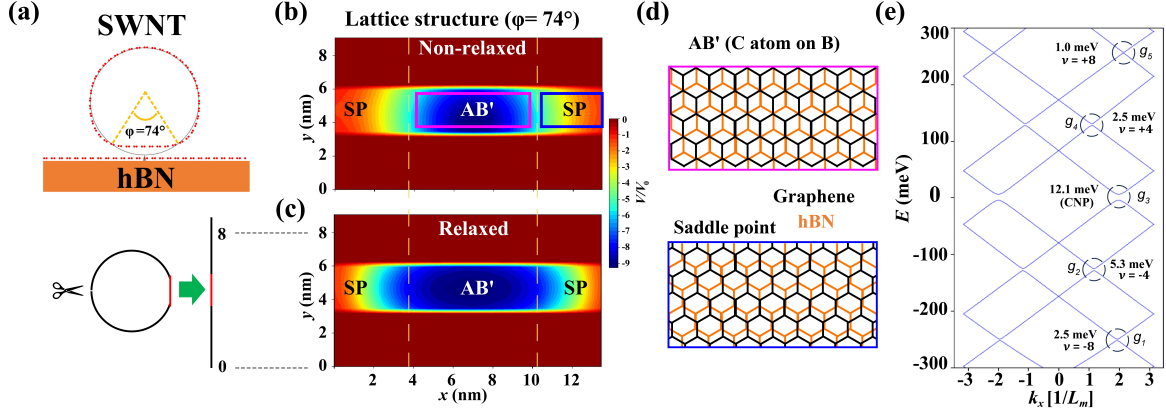

**FIG. 4. Calculated interlayer binding energy and band structure for a lattice-aligned armchair SWNT/hBN heterostructure.** (a) Top panel: DFT calculation of a commensurate SWNT on hBN for a simulation of AB domain. Partial flattening is observed with lattice relaxation (red). The flattening is characterized by the angle φ, and we find φ = 74° for the most stable configuration. Bottom panel: schematic of a "cut open" representation of interlayer binding energy in (b) and (c). (b, c) Interlayer binding energy $V$ for non-relaxed and relaxed structures, respectively, using φ = 74° and $V_0 = 0.202$ eV/nm$^2$. (d) Atomic stacking configurations for the AB' (top panel) and saddle point (SP) (bottom panel). (e) Corresponding single-particle band structure for the relaxed 1D moiré structure shown in (c), where energy gaps are clearly visible at CNP ($g_3$) and $v = \pm 4$ ($g_2$, $g_4$). The theory predicts no gapped states at $v = \pm 2$. $k_x$ is the wave vector corresponding nanotube direction and $L_m$ is the moiré length.

*Acknowledgments*—We thank Zhiwen Shi for helpful discussion. This work was supported by the National Key R&D Program of China (No. 2022YFA1203400 and No. 2023YFA1407900). M.M, R.O, N.N, M.K were supported by JSPS KAKENHI Grants (No. JP25K00938, No. JP21H05236, No. JP21H05232, No. JP24K06921) and by JST CREST Grant (No. JPMJCR20T3), Japan. N.N. also acknowledges the support from the JSPS Overseas Research Fellowship. D.K. and S.Y. were supported by the National Research Foundation of Korea (NRF) grant funded by the Korea government (MSIT) (Nos. RS-2023-00254920 and RS-2025-16067780), the Korea Institute for Advancement of Technology (KIAT) grant funded by the Korea Government (MOTIE) (No. RS00411221, HRD Program for Industrial Innovation). H.W. was supported by Sino-German Center for the Promotion of Science (GZ 1580). J.B., R.D., Z.W., S.S. and S.Z. were also supported by National Natural Science Foundation of China (No. 12174335) and Zhejiang Provincial Natural Science Foundation of China (No. LR23A040002).

# Supplementary Material

# Emergent electronic insulating states in a one-dimensional moiré superlattice

Jianfeng Bi[1]†, Masaki Minamikawa[2]†, Ruige Dong[3,1]†, DongJun Kang[4]†, Zihan Weng[1], Shaoqi Sun[1], Kenji Watanabe[5], Takashi Taniguchi[6], Ryosuke Okumura[2], Huizhen Wu[1], Naoto Nakatsuji[7]*, SeokJae Yoo[4]*, Mikito Koshino[2]*, Sihan Zhao[1]*

[1] School of Physics, Zhejiang Key Laboratory of Micro-Nano Quantum Chips and Quantum Control, Zhejiang University, and State Key Laboratory of Silicon and Advanced Semiconductor Materials, Zhejiang University, Hangzhou, China
[2] Department of Physics, The University of Osaka, Toyonaka, Osaka, Japan
[3] College of Physical Science and Technology, Dalian University, Dalian 116622, China
[4] Department of Physics, Program in Semiconductor and Device, Research and Education on Next-Generation Semiconductor Materials and Devices for Chiplet Technology, and Physics Research Institute, Inha University, Incheon, Republic of Korea
[5] National Institute for Materials Science (NIMS), 1-1 Namiki, Tsukuba 305-0044, Japan Research Center for Electronic and Optical Materials, National Institute for Materials Science, Tsukuba, Japan
[6] Research Center for Materials Nanoarchitectonics, National Institute for Materials Science, Tsukuba, Japan
[7] Department of Physics and Astronomy, Stony Brook University, Stony Brook, NY, USA

† These authors contribute equally to the work.
* Correspondence: nakatsuji0323@gmail.com; seokjaeyoo@inha.ac.kr; koshino phys.sci.osaka-u.ac.jp; sihanzhao88@zju.edu.cn

**Supplementary Note 1**

*Chirality characterization of armchair SWNTs by optical spectroscopies—*Figure 1(a) of the main text shows the Rayleigh spectrum of a suspended SWNT used for device 1, which exhibits a single prominent optical resonance at ∼ 2.18 eV ($M_{22}$ optical transition). The inset of Fig. 1(a) in the main text presents the resonant Raman spectrum of the same nanotube excited at ∼ 2.21 eV. The observed single radial breathing mode (RBM) at ∼ 101.5 cm$^{-1}$ corresponds to a SWNT with a diameter of ∼ 2.54 nm by using the empirical relation $\omega_{RBM} = 196.4 \pm 7.6/d + (24.4 \pm 3.2)$ [1]. These optical fingerprints observed in Fig. 1(a) of the main text ambiguously index the nanotube to be an armchair (18,18) [1,2]. Additionally, the rising background around the RBM peak in the inset of Fig. 1(a) in the main text at approximately 180 cm$^{-1}$ is associated with the electronic Raman scattering (ERS) feature which matches the energy difference between the excitation energy and measured $M_{22}$ optical transition [3,4], further corroborating the armchair assignment. Figure 1(c) of the main text and its inset display the Rayleigh and resonant Raman spectra of a second suspended SWNT used for device 2, respectively. The singe optical resonance at ∼ 1.88 eV ($M_{22}$ optical transition) alongside the RBM at ∼ 94.6 cm$^{-1}$ also index the nanotube to be an armchair (22,22). An ERS feature is also present near 520 cm$^{-1}$, providing further spectroscopic validation of the chirality assignment.

**Supplementary Figure 1.**

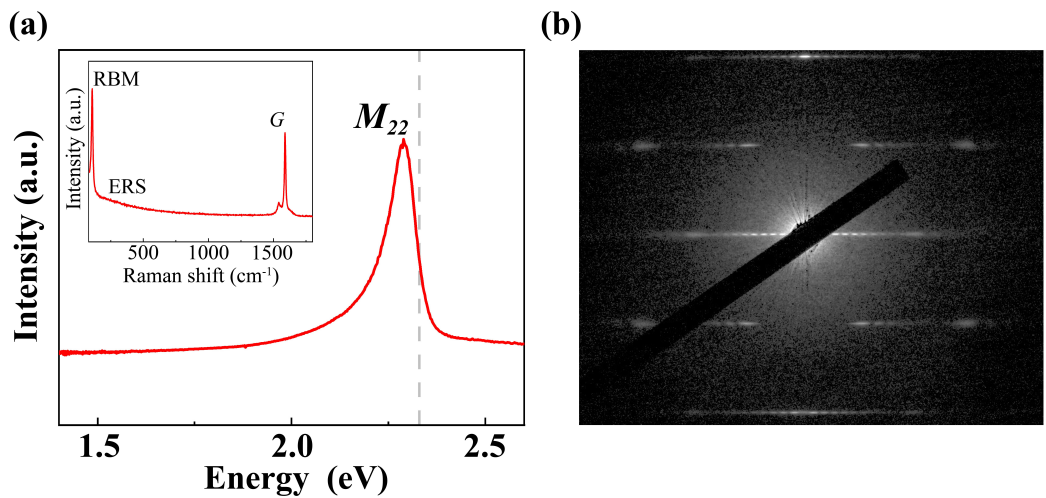

**Fig. S1. Optical spectra and corresponding nano-beam diffraction (NBD) data for an armchair SWNT (17,17).** (a) Rayleigh scattering spectrum of a suspended (17,17) SWNT exhibiting a single peak ($M_{22}$ optical transition) at ~ 2.29 eV, without spectral splitting, signifying an armchair SWNT. Inset shows the resonant Raman spectrum [laser excitation energy ~2.33 eV is indicated by the gray dashed line in (a)], where the RBM is observed at ~107.8 cm$^{-1}$, corresponding to a SWNT with a diameter of ~ 2.35 nm by using the empirical relation $\omega_{RBM}= 196.4\pm7.6/d +(24.4 \pm3.2)$ [1]. The slightly rising background around the RBM peak in the inset of (a) at approximately 300 cm$^{-1}$ is associated with the electronic Raman scattering (ERS) feature which matches the energy difference between excitation energy [dashed line in (a)] and $M_{22}$ optical transition [3,4], further corroborating the chirality assignment. (b) NBD pattern of the same SWNT shown in (a). The chirality is assigned to be (17,17) from analyzing the NBD pattern, whose result is consistent with the optical indexing shown in (a).

**Supplementary Figure 2.**

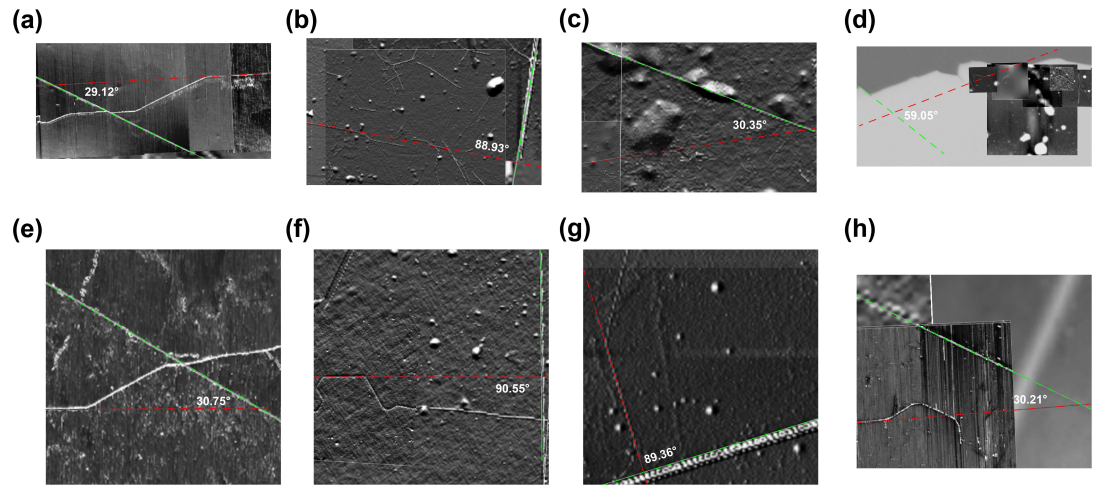

**Fig. S2. Statistical analysis of alignment angle errors between armchair SWNT and hBN.** (a-h) AFM images for eight lattice-aligned armchair SWNT/hBN heterostructures. The angles shown in (a-h) denote the measured angles between the nanotube axes (red dashed lines) and hBN's crystallographic zigzag or armchair directions of edges and/or wrinkles (green dashed lines), where 30° and its multiples are expected. The deviations from the expected 30° and its multiples are less than 1° nearly for all the samples.

**Supplementary Figure 3.**

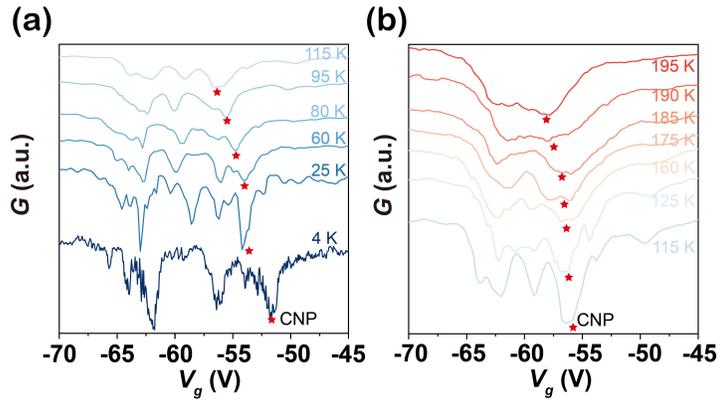

**Fig. S3. CNP identification for device 1 via gate-dependent near zero-bias conductance measurement at elevated temperatures.** By tracking the thermally driven changes of conductance with temperatures, the CNP at the lowest temperature 4K, at which Fig. 2(a) of the main text is measured, is reliably identified. Star symbols in (a) and (b) trace the CNP at different temperatures.

**Supplementary Figure 4.**

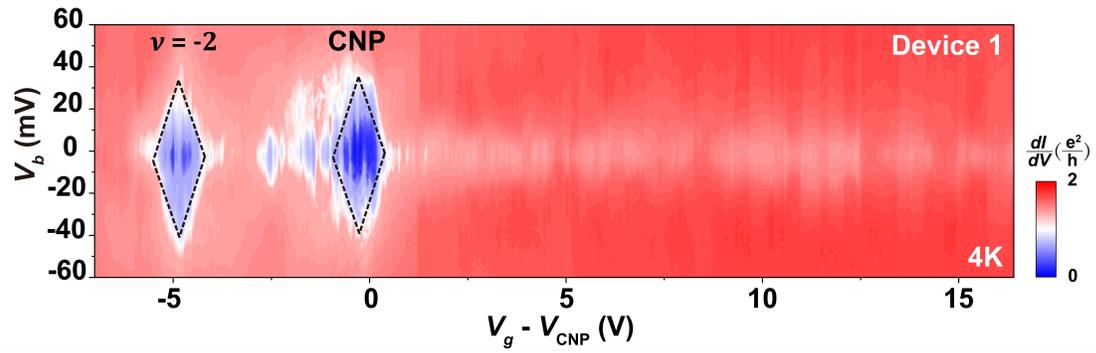

**Fig. S4. *dI/dV* map of device 1 at electron-doped side at 4 K.** We do not observe insulating behavior with pronounced diamond-like features in the electron-doped side, distinct from the hole-doped side shown in Fig. 2(a) of the main text.

**Supplementary Figure 5.**

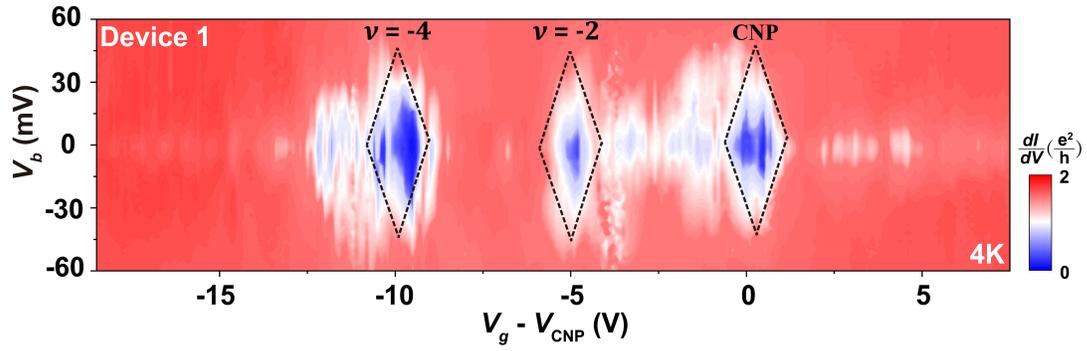

**Fig. S5. A second round *dI/dV* mapping for device 1 at 4 K.** Black dashed lines outline the pronounced diamond-like features corresponding to conductance suppressions at CNP, $v$ = -2 and -4. The data are consistent with Fig. 2(a) of the main text.

**Supplementary Figure 6.**

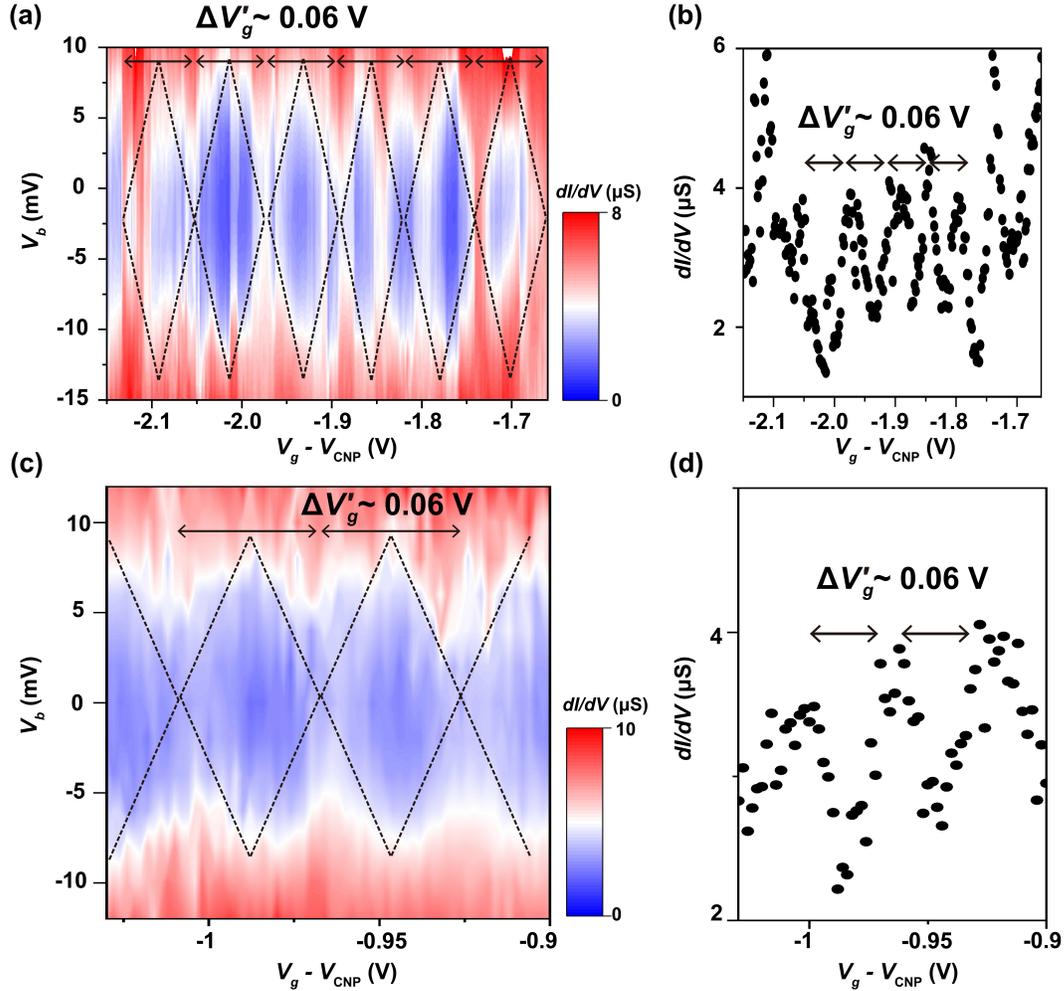

**Fig. S6. Additional data of high-resolution *dI/dV* for device 1 at 4 K.** The map is measured between the CNP and the first adjacent diamond ($v = -2$) with a gate voltage step of 2 mV. Our device is made of a metallic armchair SWNT, and the Coulomb blockade (CB) is not very strong. (a) High-resolution *dI/dV* for device 1. The Coulomb diamonds reveal CB behavior with approximately a peak spacing of ~ 0.06 V as indicated by the black dashed lines, from which we estimate a gate capacitance of $C_g$ ~ 5.3 pF/m. (b) Near zero-bias line cut of the *dI/dV* map shown in (a), where an approximate peak spacing of $\Delta V_g^{'}$ ~ 0.06 V can be discerned (indicated by the black double arrow). (c) Repeated measurement in a different measurement round. The diamond-like features showing CB behavior are indicated by the black dashed lines and $\Delta V_g^{'}$ ~ 0.06 V is also visible, albeit it is not as clear as that shown in (a) after long time *dI/dV* scans. (d) Near zero-bias line cut corresponding to the data in (c). An approximate peak spacing of $\Delta V_g^{'}$ ~ 0.06 V can also be discerned (indicated by the black double arrows)

**Supplementary Note2.**

*COMSOL simulation details—*We numerically calculate the electrostatically induced charge density and the geometric capacitance of SWNTs using AC/DC module of COMSOL Multiphysics. A $SiO_2$ insulator has the relative permittivity $\varepsilon_{SiO2}=3.9$, while hBN has the anisotropic permittivity of $\varepsilon_{/\!/} = 3.5$ and $\varepsilon_\perp = 6.9$ along the in-plane and out-of-plane directions, respectively. A Si back-gate is modeled as a thin 2D metal applied by the gate voltage $V_g$. SWNT is modeled by a solid cylinder of length $L$ and radius $R$. SWNT radius $R$ is determined by its chirality index.

To calculate the electrostatic response of SWNTs, we use two different models: (i) perfect electric conductor (PEC) model and (ii) finite density of state (DOS) model. The former takes account only with the geometric capacitance $C_g$ by assuming infinite number of charge carriers, while the latter supposes finite number of charge carriers due to finite DOS, thus including both effects from the geometric capacitance $C_g$ and the quantum capacitance $C_Q$.

In the PEC model, the charges are induced only on the cylinder surface upon the gate voltage $V_g$. Then, the total charge Q gate-induced in SWNT is obtained by the surface-integration, $Q = \oint_{SWNT} \sigma(\mathbf{r}) da$, where $\sigma(\mathbf{r})$ denotes the surface charge density. Then, the geometric capacitance $C_g$ is given by $C_g = Q/V_g$, while other conductive elements such as source and drain electrodes are grounded. We also convert the surface charge density into the line charge density by averaging $\sigma(\mathbf{r})$ over its circumference, i.e. $p(z) = \int_0^{2\pi} \sigma(\mathbf{r}) R d\phi / e$, where $z$ is the direction along the SWNT axis. We refer to $p(z)$ as the hole density hereinafter since we calculate the charges upon negative gate voltage $V_g < 0$.

In the finite DOS model, the following tight-binding model expressions are employed to describe the DOS of armchair SWNTs [5],

$$DOS(E) = \sum_{q=1}^{2n} g(E,q), \quad (1)$$

where the subband DOS is given by

$$g_{ac}(E,q) = \frac{2\alpha}{a\pi^2 R} \frac{|E|}{\sqrt{E^2 - E_{vh1}^2} \sqrt{\left(\sqrt{E^2 - E_{vh1}^2} - A_{c1}\right)\left(-\sqrt{E^2 - E_{vh1}^2} + A_{c2}\right)}}. \quad (2)$$

Here, $n$, $q$, $a$, and $\alpha$ are the chirality index of SWNT, subband index, graphene lattice constant (~2.46 Å) and zone degeneracy (1 at Γ-point and 2 otherwise), respectively. Note that the factor ($2\pi R$) is introduced in Eq. (2) to match DOS unit with our 3D numerical simulation. $E_{vh1} = \pm |\gamma_0 \sin(\pi q / n)|$ represents the van Hove singularities of each subband, while $A_{c1} = \gamma_0 \left[ -2 + \cos(\pi q / n) \right]$ and $A_{c2} = \gamma_0 \left[ 2 + \cos(\pi q / n) \right]$ are armchair energy parameters. The nearest-neighbor overlap energy $\gamma_0$ is set to 2.7 eV, which lies within the range of nominal range (2.5-3.2 eV) [5]. The local hole density is obtained by the integration of DOS, Eq. (2),

$$\sigma(\mathbf{r}) = e \sum_{q=1}^{2n} \int_{E_{\min}(q)}^{E_{\max}(q)} g(E,q) \left[ 1 - \frac{1}{e^{(E-E_F-V(\mathbf{r}))/k_B T}+1} \right] dE, \qquad (3)$$

taking the hole occupation into account. In DOS model, $\sigma(\mathbf{r})$ is a function of electrostatic potential, and thus calculations of $\sigma(\mathbf{r})$ (Eq. (3)) and potential ($V$) are needed to be self-consistently integrated in COMSOL Multiphysics. Then, the line charge density is calculated by the same manner, i.e. $p(z) = \int_0^{2\pi} \sigma(\mathbf{r}) R d\phi / e$, as in PEC model.

To determine the filling fraction $v$, we estimate the device capacitance through electrostatic simulations of the geometric capacitance $C_g$. Fig. S7(a) shows equipotential contours for device 1, which consists of the back-gate, $SiO_2$, hBN, two electrodes, and an (18, 18) SWNT when $V_g = -5$ V is applied to the back-gate. The SWNT is modeled as a perfectly conducting metallic cylinder (i.e. PEC model) whose radius is determined by its chirality index (18, 18). As shown in Fig. S7(a), the large Au electrodes (2 μm-wide and 50 nm-thick) significantly screen the potential from the back-gate. This screening effect results in a suppressed geometric capacitance of $C_g = 9.93$ pF/m. The equipotential contours in Fig. S7(b) demonstrates that the potential by the back-gate efficiently affect SWNT, resulting in relatively high charge density on the SWNT surface. We find that the electrodes (S and D) of device 1 in Fig. S7(a) suppress the capacitance by 43.42% from $C_g = 22.87$ pF/m [Fig. S7(b)] to 9.93 pF/m [Fig. S7(a)]. Important to note that the calculated 9.93 pF/m is the higher bound for the gate capacitance of this device, the actual value should be smaller than it due to the unavoidable charged traps and impurities. Insets in Fig. S7(a)&(b) show the different line charge accumulation upon the same gate voltage $V_g = -5$ V. We also include the finite density of states of SWNTs (i.e. DOS model) in our calculation [5], and the result demonstrates the quantum capacitance effect is marginal, i.e. the total capacitance $C_g = 9.21$ pF/m, ~7% smaller than $C_g = 9.93$ pF/m in a metallic cylinder model (i.e. PEC model). The equipotential contour plots of the SWNT devices with a finite density of states are nearly identical to those without it.

For (22,22) SWNT device (device 2), we obtain $C_g = 9.55$ pF/m [Fig. S10(a)], while the geometry without electrodes yields $C_g = 24.70$ pF/m [Fig. S10(b)]. The effect of the capacitance suppression by the electrodes is 38.66%.

**Supplementary Figure 7.**

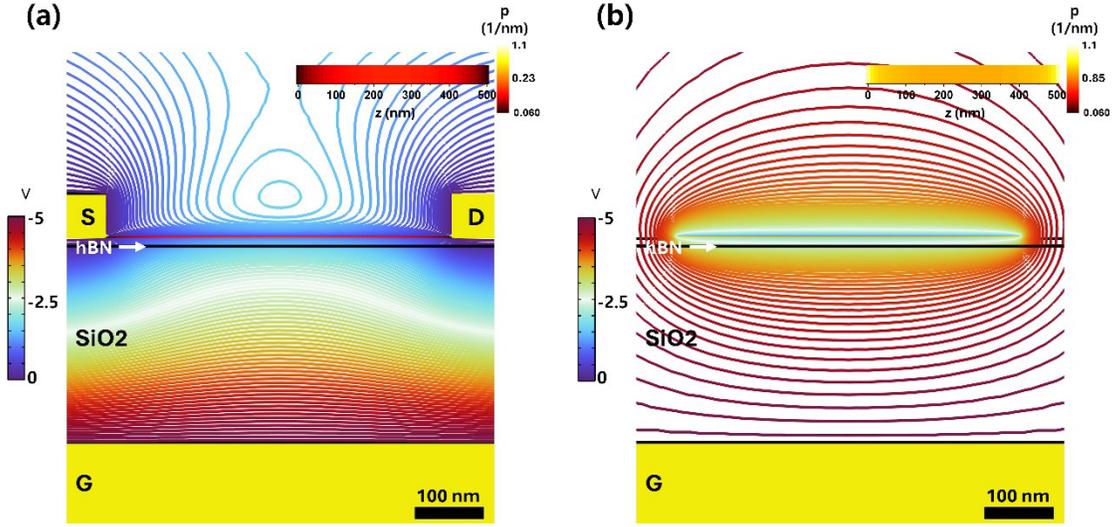

**Fig. S7. Electrostatic calculation for (18, 18) SWNT device**. (18,18) SWNT is positioned on a hBN layer (10 nm-thick). A Si back gate (G) is separated from the hBN/SWNT stack by a SiO$_2$ dielectric layer (285 nm-thick). The rainbow-colored equipotential contours demonstrate that the SWNT is directly exposed to the electric potential induced by gate bias without electrode screening. The inset shows the hole density of (18,18) SWNT, $p(z)$, evaluated at every 10 nm interval, using a black-to-white color scheme. (a) (18, 18) SWNT device with electrodes. The equipotential contours demonstrate the potential screening by the electrodes (S and D), which leads to a relatively small charge density on the surface of SWNT (inset), resulting in effective suppression of SWNT capacitance. (b) (18,18) SWNT device without electrodes. The equipotential contours demonstrate that the SWNT is directly exposed to the electric potential induced by the gate voltage without the electrode screening, resulting in a relatively large charge density on the surface of SWNT (inset).

For this device, by taking into account the finite density of states of the SWNT, we find that the quantum capacitance slightly reduces the overall capacitance, resulting in a total capacitance of $C_g \sim 9.21$ pF/m. Important to note that the calculated 9.21 pF/m is the higher bound for the gate capacitance of this device, the actual value should be smaller than it due to the unavoidable charged traps and impurities.

**Supplementary Figure 8.**

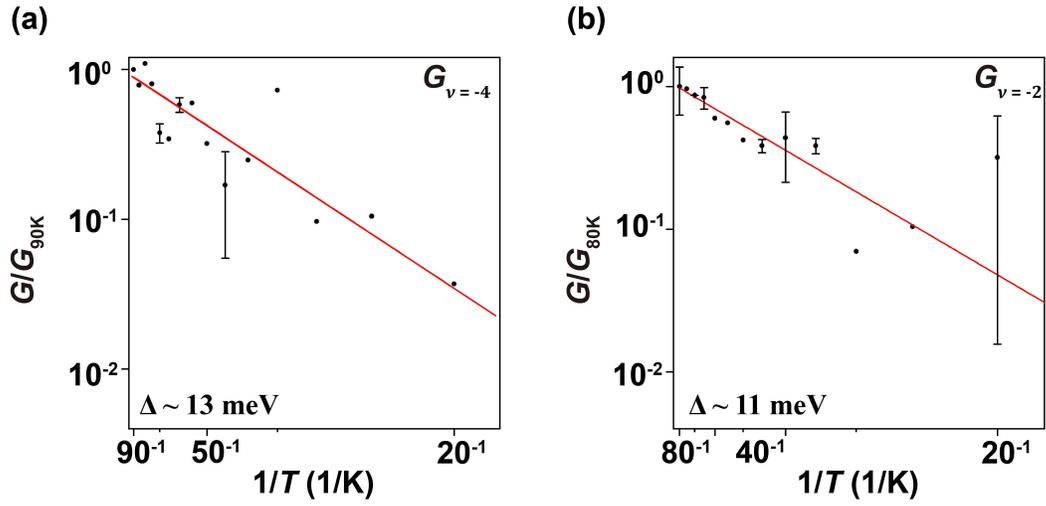

**Fig. S8. Arrhenius analysis at full and half moiré filling for device 1.** (a, b) Renormalized near-zero bias conductance at full filling [($v$ = -4 in (a)] and half moiré fillings [($v$ = -2 in (b)]. The conductance in (a) and (b) is normalized to that measured at 90 K and 80 K, respectively. Black symbols show the mean values at each temperature and error bars indicate the standard deviations. The red lines in (a) and (b) are the fits for a thermally activated model with the form $G_{CNP}(T) \propto \exp(-\Delta/2T)$, where $\Delta$ is the energy gap. The fitting yields an energy gap of $\Delta \sim 13$ meV [($v$ = -4 in (a)] and $\Delta \sim 11$ meV [($v$ = -2 in (b)], respectively.

**Supplementary Figure 9.**

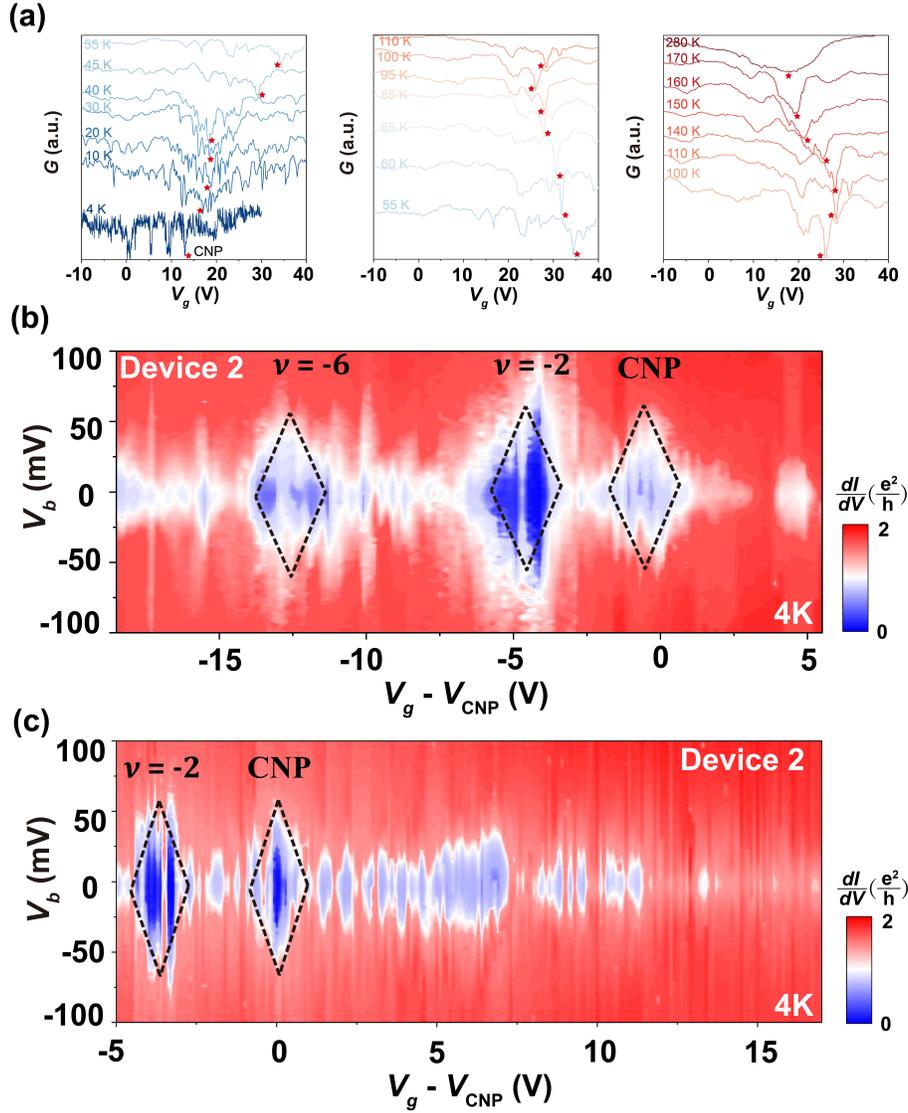

**Fig. S9. A second round *dI/dV* mapping on the hole-side, data on electron-side, and details on CNP identification for device 2.** (a) CNP identification for device 2 via gate-dependent near zero-bias conductance measurement at elevated temperatures. By tracking the thermally driven changes of conductance with temperatures, the CNP at the lowest temperature 4K, at which Fig. 2(e) of the main text is measured, is identified. Star symbols in (c) trace the CNP at different temperatures. (b) A second round *dI/dV* mapping for device 2 after Fig. 2(e) of the main text. Black dashed lines outline the large diamond-like features corresponding to conductance suppressions at CNP, $v = -2$, and $-6$. We note that $v = -4$ insulating features is not clearly visible in the second-round measurement. (c) Data for electron-doped region measured concurrently with Fig. 2(e) of the main text.

**Supplementary Figure 10.**

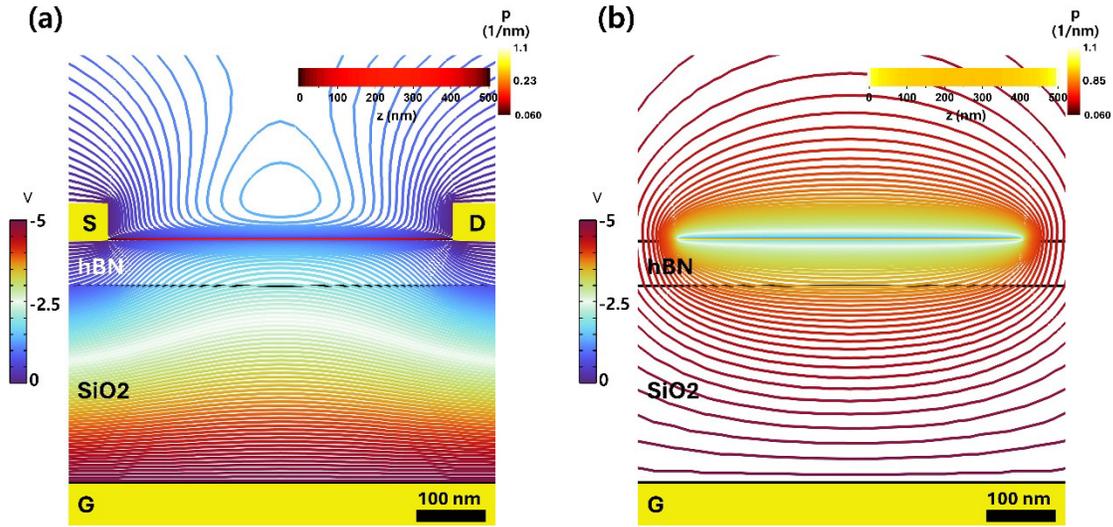

**Fig. S10. Electrostatic calculation for (22,22) SWNT device.** (22,22) SWNT is positioned on a hBN layer (65 nm thick). A Si back gate (G) is separated from the hBN/SWNT stack by a $SiO_2$ dielectric layer (285 nm thick). The rainbow-colored legend indicates the potential of the equipotential contours. The inset shows the hole density of (22,22) SWNT, evaluated at every 10 nm interval, using a black-to-white color scheme. (a) (22,22) SWNT device with electrodes. The equipotential contours demonstrate the potential screening by the electrodes (S and D), which leads to a relatively small charge density on the surface of SWNT (inset), resulting in effective suppression of SWNT capacitance. (b) (22,22) SWNT device without electrodes. The equipotential contours demonstrate that the SWNT is directly exposed to the electric potential induced by the gate voltage without the electrode screening, resulting in a relatively large charge density on the surface of SWNT (inset).

For this device, we obtain $C_g \sim 9.55$ pF/m [Fig. S10(a)], which closely matches that of device 1 [Fig. S7(a)].

**Supplementary Figure 11.**

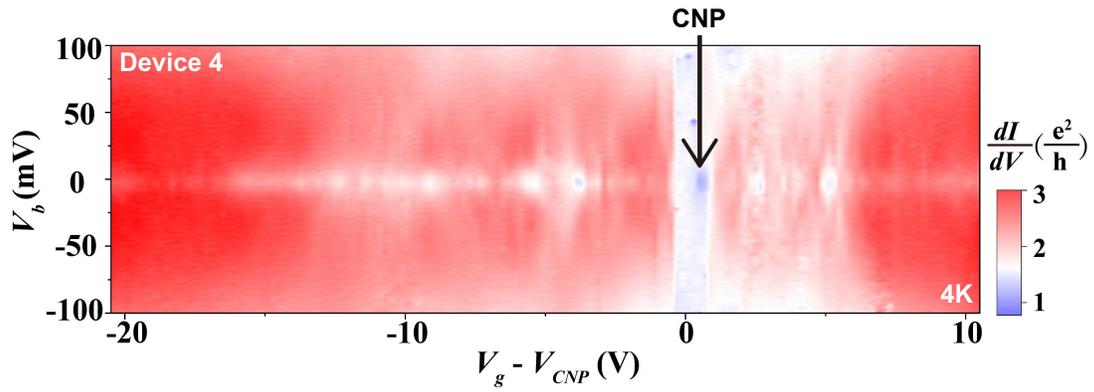

**Fig. S11. *dI/dV* color plot for an extra armchair (16,16) SWNT device on SiO₂/Si, without hBN alignment (device 4).** The data are measured at 4 K. Unlike devices with hBN alignment, no clear pronounced diamond-like insulating features are visible.

**Supplementary Note 3**

*Model for SWNT on hBN—*We consider an armchair $(n,n)$ SWNT on monolayer hBN. The $x$ axis is taken along the tube axis and the $z$ axis is perpendicular to the hBN plane. The origin is defined at the point where the SWNT and hBN are closest, with an interlayer distance $d_0 = 0.322$ nm [6]. The chiral vector is $C_h = n(a_1+a_2)$, and the translational vector is $T = |a_j|R(90°)C_h/|C_h|$, where $a_j$ denotes the primitive lattice vector of an unfolded graphene sheet. The hBN layer is aligned with the SWNT so that both share the zigzag direction along $x$. The moiré period due to the lattice mismatch is $L=(1+1/\epsilon)a_G$, with $1+\epsilon=a_{hBN}/a_G \approx 1.018$ [7], where G is graphene, $a_G=0.246$ nm and $a_{hBN}=0.2505$ nm are the lattice constants of graphene and hBN, respectively.

As shown by commensurate-approximation DFT calculations [see Fig. 4 (a) in the main text and Supplementary note 5], the SWNT adopts a 90°-rotated D-shaped geometry. We define the flat region length as $l_\phi=2R_t\sin\phi$, where $R_t$ is the tube radius. In the following, we assume that only carbon atoms within this region interact with the hBN layer, as the distance outside it increases rapidly.

In the real system, moiré-induced lattice relaxation occurs. Because direct DFT calculations for the moiré structure are computationally demanding, we employ a spring–mass approach based on the continuum model for moiré bilayers [Supplementary note 4] [8]. In this model, the hBN layer is allowed to relax within the $xy$ plane, while the SWNT relaxes while keeping the overall 90°-rotated D-shaped geometry fixed. We further assume that only the flat region within $|y| \leq l_\phi$ interacts with hBN through the effective interlayer binding energy

$$U_B = \sum_{i \in \text{overlap}} [\theta(Y_i - l_\phi/2) - \theta(Y_i + l_\phi/2)] V_B(\mathbf{R}_i) l_0^2, \quad (4)$$

where $V_B(\mathbf{R}_i)$ is the local binding energy of G/hBN [defined later in Eq. (8)], and $\theta(x)$ is the step function.

The electronic properties are studied using a tight-binding model for G/hBN [7,8]. To construct a commensurate supercell, we take $a_{hBN}/a_G=56/55$. The total Hamiltonian is written as $H=H_G+H_{hBN}+H_T$, where

$$H_G = -\sum_{i \neq j} t_G(\mathbf{R}_i - \mathbf{R}_j) c_i^\dagger c_j, \quad H_{hBN} = \sum_i V_i d_i^\dagger d_i,$$
$$H_T = -\sum_{i,j} t_{G\text{-}BN}(\mathbf{R}_i, \mathbf{R}_j) c_i^\dagger d_j + \text{h.c.} \quad (5)$$

The hopping functions are defined as $t_G(\mathbf{R}_i, \mathbf{R}_j) = t(\mathbf{R}_i - \mathbf{R}_j)$ and $t_{G\text{-}BN}(\mathbf{R}, \mathbf{R}') = [\theta(Y - l_\phi/2) - \theta(Y + l_\phi/2)] t(\mathbf{R}' - \mathbf{R})$, where the interlayer hopping is assumed to occur only in the flat region. The hopping amplitude $t(\mathbf{R})$ follows the Slater–Koster form,

$$-t(\mathbf{R}) = V_{pp\pi} \left[1 - \left(\frac{\mathbf{R} \cdot \mathbf{e}_z}{R}\right)^2\right] + V_{pp\sigma} \left(\frac{\mathbf{R} \cdot \mathbf{e}_z}{R}\right)^2,$$
$$V_{pp\pi} = V_{pp\pi}^0 e^{-(R-a/\sqrt{3})/r_0}, \quad V_{pp\sigma} = V_{pp\sigma}^0 e^{-(R-d)/r_0}, \quad (6)$$

where $\mathbf{e}_z$ is the unit vector along $z$ axis, and we adopt $V_{pp\pi}^0 \approx -2.7$ eV, $V_{pp\sigma}^0 \approx 0.48$ eV, and $r_0=0.184a_G$ [9]. The $V_i$ is the on-site potential on hBN layer, which is set as $V_B = 3.34$ eV and $V_B = -1.40$ eV [9].

**Supplementary Note 4**

*Spring-mass model*—We briefly review the spring–mass model for G/hBN (see [8] for details). Each layer is represented by a square grid of masses spaced by $l_0$, connected by springs along orthogonal and diagonal directions. The grid spacing $l_0$ is chosen to be much smaller than the moiré length scale, (here we use $l_0 = 1.27 a_G$). The total energy $U = U_E + U_B$ is minimized with respect to the in-plane displacement field $u(r_i)$ at the mass points $r_i$. The elastic energy is given by

$$U_E = \sum_{l=1,2} \left[ \frac{k^{(l)}}{2} \sum_{\langle i,j \rangle} \delta d_{ij}^{(l)2} + \frac{k_d^{(l)}}{2} \sum_{\langle\langle i,j \rangle\rangle} \delta d_{ij}^{(l)2} + \frac{\kappa^{(l)}}{2\ell_0^2} \sum_i \left( S_i^{(l)} - \ell_0^2 \right)^2 \right], \qquad (7)$$

where $\langle i,j \rangle$ and $\langle\langle i,j \rangle\rangle$ denote nearest and diagonal pairs, $\delta d_{ij}^{(l)} = |\mathbf{r}_i + \mathbf{u}_i^{(l)} - \mathbf{r}_j - \mathbf{u}_j^{(l)}| - |\mathbf{r}_i - \mathbf{r}_j|$, and $S_i^{(l)}$ is the area of the deformed plaquette. The force constants are $k^{(1)} = 15.6$ eV/Å², $k_d^{(1)} = 7.3$ eV/Å², $\kappa^{(1)} = -4.3$ eV/Å² for hBN, and $k^{(2)} = 19.14$ eV/Å², $k_d^{(2)} = 9.57$ eV/Å², and $\kappa^{(2)} = -6.32$ eV/Å² for graphene. The interlayer binding energy is

$$\begin{aligned} U_B^{(G/BN)} &= \sum_{i \in \text{overlap}} V_B(\mathbf{r}_i) \ell_0^2, \\ V_B(\mathbf{r}) &= V_1 + 2V_0 \sum_{\alpha=1}^3 \cos\left[ \mathbf{G}_\alpha^M \cdot \mathbf{r} + \bar{\mathbf{b}}_\alpha \cdot \left( \mathbf{u}^{(1)} - \mathbf{u}^{(2)} \right) + \varphi_0 \right], \end{aligned} \qquad (8)$$

where $\mathbf{G}_\alpha^M = \mathbf{b}_\alpha - \tilde{\mathbf{b}}_\alpha$ are the moiré reciprocal vectors of graphene/hBN, and $\bar{\mathbf{b}}_\alpha = (\mathbf{b}_\alpha + \tilde{\mathbf{b}}_\alpha)/2$. $\mathbf{b}_\alpha$ and $\tilde{\mathbf{b}}_\alpha$ are the reciprocal lattice vector of the graphene and hBN layers, respectively. The parameters are given by $V_0 = 0.202$ eV/nm², $V_1 = -0.700$ eV/nm², and $\phi_0 = 0.956$ [10,11].

**Supplementary Note 5**

*Commensurate approximation and DFT calculation*—To avoid the large moiré supercell, we consider a commensurate SWNT/hBN system with matched lattice constants ($a_G = a_{hBN}$). DFT calculations were performed using Quantum Espresso [12] with the PBE–PAW functional and DFT-D3 correction for van der Waals interactions. 12×1×1 k-point mesh was used. For the y direction, hBN atoms within a cutoff $l_{cutoff} = 2R_t$ were included, which is sufficient for interlayer decoupling. We performed vc-relax calculations with different initial interlayer distances — 50%, 30%, 20%, and 10% of the interlayer spacing of bilayer graphene 0.335 nm — and selected the most stable structure with $\phi = 74°$, corresponding the initial condition 20%. The hBN layer was fixed, while SWNT atoms were relaxed until the force on each atom was less than 0.005 eV/Å.

**Supplementary Figure 12.**

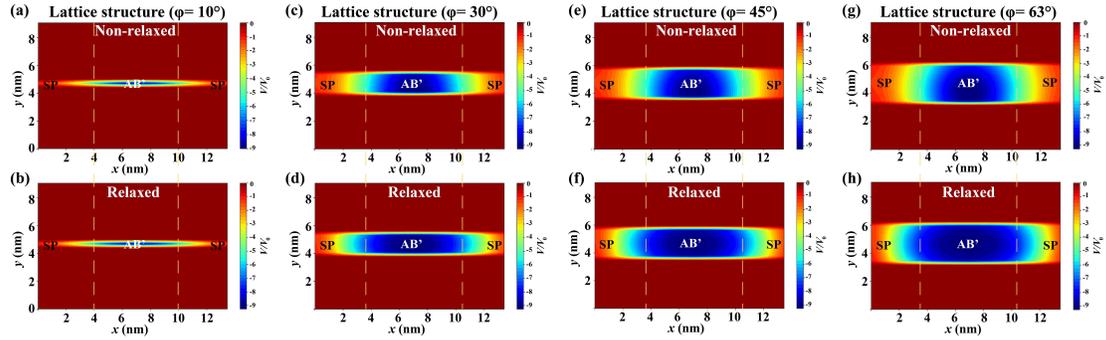

**Fig. S12. Calculated interlayer binding energy at different φ for an armchair SWNT/hBN heterostructure.** Projected interlayer binding energy landscape for non-relaxed and relaxed lattices, respectively, at φ = 10° (a, b); φ = 30° (c, d); φ = 45° (e, f); φ = 63° (g, h). The x and y are along the tube axial and circumferential directions, respectively. Two relevant stacking configurations for our heterostructures are AB' and saddle point (SP). AB' stacking is most energetically favorable where carbon atoms of the SWNT are aligned directly over boron atoms in the hBN; while the SP configuration corresponds to a transitional region that is energetically unfavorable. Compared to the non-relaxed case, the interlayer binding energy landscape for the relaxed case clearly shows the expansion of AB' (in both x and y directions) at the cost of SP area for all different φ (eye guided by the yellow dashed lines). The diameter of the armchair SWNT used for the calculation is about 3 nm, approximating those used in devices 1 and 2 of the main text.

**Supplementary Figure 13.**

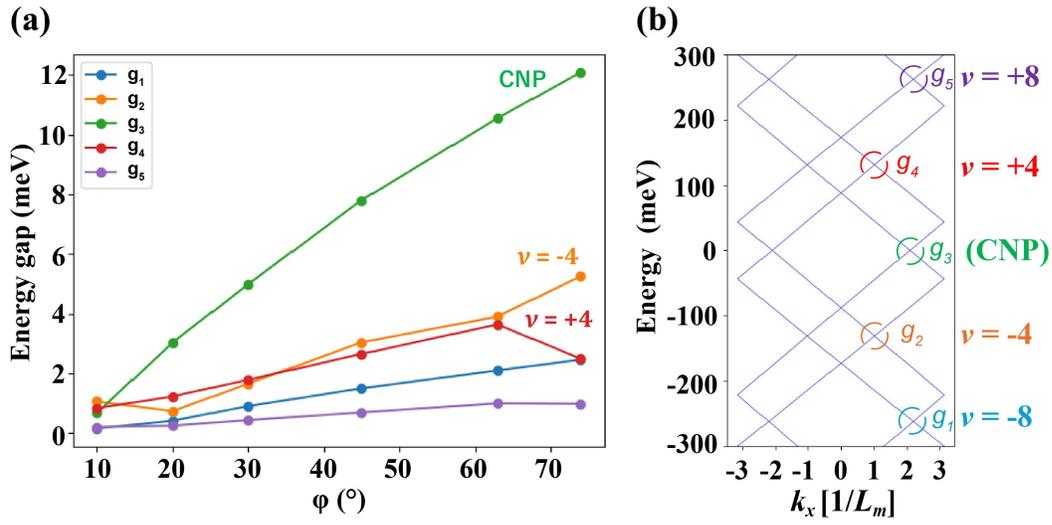

**Fig. S13. Single-particle band gap evolution on φ for an armchair SWNT/hBN heterostructure.** (a) Evolution of the band gaps as a function of φ for an armchair SWNT/hBN heterostructure used in Fig. 4 of the main text. Five characteristic gaps are highlighted [also see (b)]: $g_3$ at the CNP (green), $g_2$ at $v = -4$ (orange), $g_4$ at $v = +4$ (red), $g_1$ at $v = -8$ (blue), and $g_5$ at $v = +8$ (purple). (b) Calculated band structure of the armchair SWNT without hBN potential, illustrating the relevant 'gaps' in (a) from $g_1$ to $g_5$. $k_x$ is the wave vector corresponding nanotube direction and $L_m$ is the moiré length.